\documentclass[sigconf, nonacm]{acmart}

\renewcommand\footnotetextcopyrightpermission[1]{} % removes footnote with conference information in first column
\AtBeginDocument{%
  \providecommand\BibTeX{{%
    \normalfont B\kern-0.5em{\scshape i\kern-0.25em b}\kern-0.8em\TeX}}}

\setcopyright{none}

\usepackage{amsmath,amssymb,amsfonts}

\usepackage{textcomp}
\usepackage{xcolor}
\usepackage[caption=false,font=footnotesize]{subfig}
\usepackage{booktabs} % For formal tables

\graphicspath{{pdf/}{jpeg/}{png/}{eps/}{Images/}}
\usepackage{stfloats}
\usepackage{balance}
\usepackage{soul}
\usepackage[normalem]{ulem}
\usepackage{cancel}
\usepackage{verbatim}
\usepackage{graphicx}
\usepackage{cases}

\usepackage{listings}

\begin{document}

\title{Utilizing Opportunistic Social Networks for Remote Patient Monitoring in Rural Areas}

\author{Esther Max-Onakpoya}
\affiliation{%
  \institution{Department of Computer Science\\ 
        University of Kentucky, Lexington\\
        esther.max05@uky.edu
        }
  }
\author{Shina Madamori}
\affiliation{%
  \institution{Department of Computer Science\\ 
        University of Kentucky, Lexington\\
        shina@uky.edu
        }
  }
\author{Corey E. Baker}
\affiliation{\institution{Department of Computer Science\\ 
        University of Kentucky}}
\email{baker@cs.uky.edu}

\renewcommand{\shortauthors}{E. Max-Onakpoya et al.}

\begin{abstract}
  The use of Internet connectivity for remote patient monitoring is often unsuitable 
for rural communities where Internet infrastructure is lacking, and power outages are 
frequent. 
This paper explores the rural connectivity problem in the context of remote patient 
monitoring and analyzes the feasibility of utilizing a delay tolerant network (DTN) architecture 
that leverages the social behaviors of rural community members to enable out-of-range 
monitoring of patients in rural communities without local transportation systems.
The feasibility is characterized using delivery latency and delivery rate with the number of 
participants and the number of sources as variables.
The architecture is evaluated for Owingsville, KY using U.S. Census Bureau, the National Cancer Institute's, and IPUMS ATUS sample 
data.
The findings show that within a 24 hour window, there is an exponential relationship 
between the number of participants in the network and the delivery rate with a minimal 
delivery of 38.7\%, a maximal delivery rate of a 100\% and an overall average delivery rate of 89.8\%.
\end{abstract}










\maketitle

\section{Introduction}\label{sec_intro}
\begin{figure*}[htbp]
\centering
   \includegraphics[width=0.75\textwidth, height=3.0in]{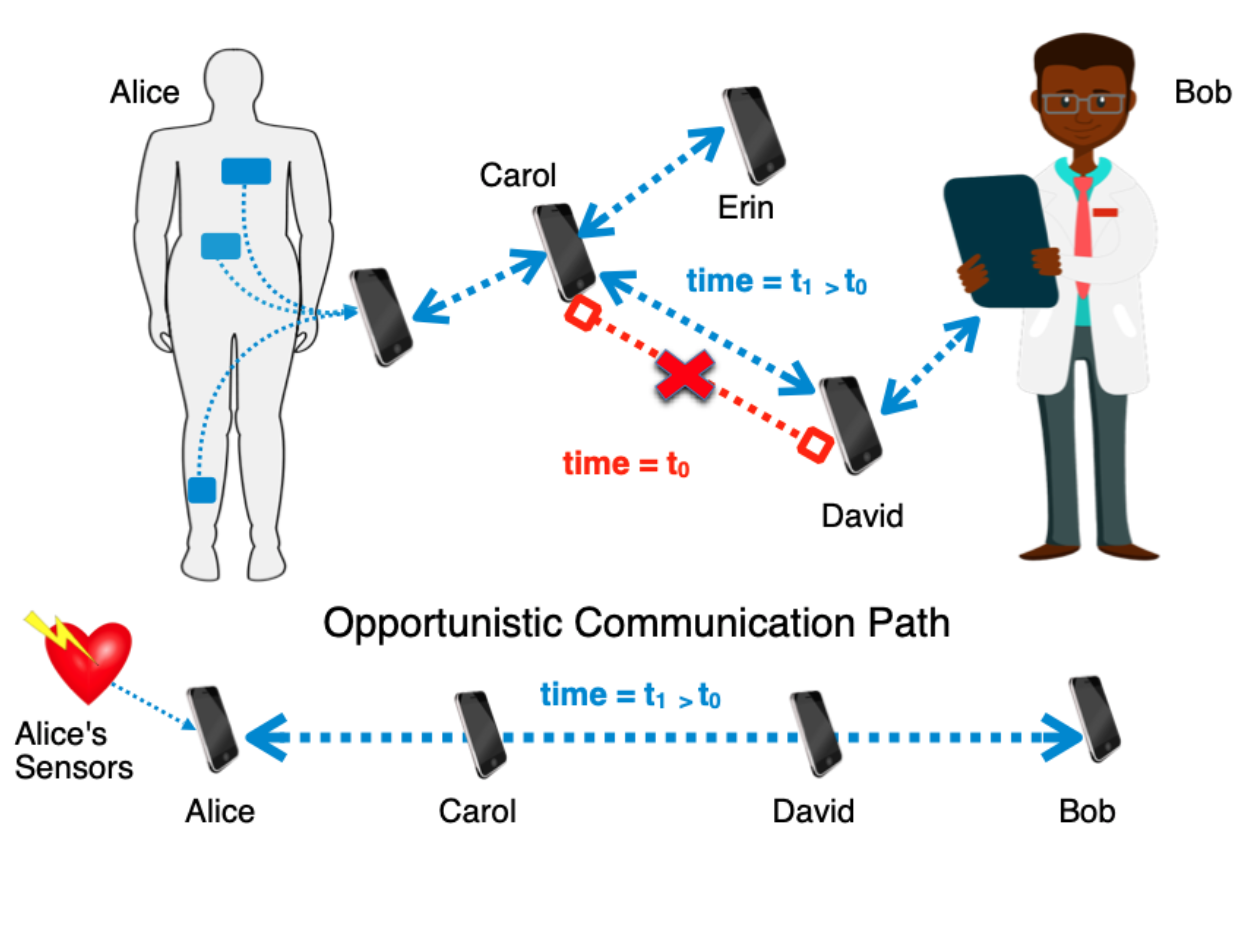}
\caption{Delay Tolerant Network Path for Remote Patient Monitoring}
\label{fig1}
\end{figure*}

The ubiquity of mobile devices and rapid improvement in wireless body sensors has revolutionized the field of healthcare.
Through mHealth solutions, practitioners can remotely monitor and assist with patients' disease management in real time or asynchronously.
This has improved the timeliness of clinical decision making, decreased the length of hospital stays, and reduced mortality rates~\cite{niksch2014value,moy2017leading}.
Although many patients have benefited from mHealth solutions, and national efforts are underway to accelerate broadband deployment in under-served areas of the US~\cite{usda2019}, rural patients may not benefit to the same extent as their non-rural counterparts due to geographical and financial barriers that result in limited or nonexistent access to broadband connectivity~\cite{henley2017invasive}.
Unfortunately, rural patients have both a higher prevalence of chronic disease and lower access to medical care~\cite{services2017}.
A promising solution lies in the use of a delay tolerant network (DTN) architecture that leverages human mobility for disseminating patient health information (PHI) to care entities (Figure \ref{fig1}).

%-----------------------------------------------------------------------------
Though there has been some important research conducted to connect rural areas 
using DTNs~\cite{pentland2004daknet,doria2002providing,whitbeck2010hymad,raffelsberger2013hybrid}, the network characteristics
such as node mobility and density differ vastly resulting in the need for a better 
understanding of how DTNs will operate in rural remote patient monitoring (RRPM) scenarios.
A survey conducted in~\cite{syed2012study} that consisted of clinical and non-clinical staff 
showed that DTNs are believed to be a promising solution for remote patient monitoring in 
low-resource settings for number of application domains including EHRs, notifications, 
new \& blogs, etc.
Additionally, Syed et. al. reports that there has been 
limited work on evaluating how DTNs will perform in RRPM environments~\cite{syed2012study}.
Understanding how the use of human mobility influences data dissemination 
in RRPM environments is essential, but can be cumbersome due to its inherent characteristics.

Unlike other networks, such as social networks, in which most of the 
population is able to actively participate, patient monitoring networks typically consist of a small 
percentage of the population: patients, caregivers, and healthcare 
providers.
In areas without public transportation, DTNs for the RRPM case will have to solely depend on opportunistic encounters between people.
Hence, this work (1) derives a simple and novel mathematical description for delay tolerant remote monitoring in rural communities; and
(2) provides insights on how the intrinsic characteristics of a real rural city, Owingsville, KY,
influences network performance of delay tolerant rural patient data when the dissemination of 
information depends solely on inherent community participation.

\section{Related Work}\label{section:related}

%-----------------------------------------------------------------------------
Previous work have explored various means of reducing the disparity in rural 
connectivity compared to their urban counterparts~\cite{doria2002providing,pentland2004daknet,whitbeck2010hymad,raffelsberger2013hybrid}.
In the area of RRPM, some mobile ad-hoc network schemes have been proposed for 
transmitting PHI between patients and medical staff within a hospital~\cite{rashid2005real,cho2008opportunistic}.
Other researchers have investigated propagating voicemails or other emergency 
data via DTNs in amongst communities~\cite{adlassnig2009rural,jang2009rescue}. 
More closely related work proposes using vehicular ad hoc networks to disseminate 
PHI in rural communities~\cite{barua2014rcare,murillo2011application}.

The primary limitation of the aforementioned work is the lack of evidence that 
evaluations adequately reflect real rural environments, particularly when it 
comes to mobility and density; which 
are all required to properly assess solutions in rural environments. 
Additionally, the aforementioned work solely utilize vehicular communications 
in transmitting messages to medical entities.
In regions where periodic vehicular networks (such as bus systems, etc) do not exist, 
the DTN would have to depend on other forms of node mobility~\cite{madamori2019using}.
%-----------------------------------------------------------------------------
This work is unique because it utilizes data from a real rural city, Owingsville, 
KY to obtain the natural mobility and density of nodes, and does not rely on local transportation systems.

\section{Modeling Rural Remote Patient Monitoring}\label{section:real_world}
In order to understand the viability and feasibility of the network, a simple mathematical model is introduced that depicts some real-world characteristics of the RRPM problem, including the low-density of source-nodes(patients) and limited number of points of interests (POIs).

\subsection{Model description}\label{model_desc}
Consider a network in which $N$ represents a set of nodes that are randomly distributed in a square grid divided into $M \times M$ cells, where an individual node is represented by $n \in N$. 
Let $\{A,R,P,D\} \subseteq N$
, where subsets are defined as: $A$ - patients, $R$ - relay nodes, $P$ - points of interest and $D$ - destinations,
such that: 

\begin{equation}
    |D| \leq |A| << |R| 
\end{equation}\label{equation:nodes}

\noindent Let the set $\{E,U\}\subseteq R$ represent employed and unemployed intermediary nodes respectively.
Let an individual patient be represented by $a \in A$ and the set of messages for an individual patient be:

\begin{equation}
    m_{a,j}(t_{0}) \in m_a \mid 0 < j \leq |m_a|, t_{0} = t
\end{equation}

\noindent where message number $j$ is generated at time $t_0$ and is transmitted to $D$ at a time $t \leq t_{f}$, where, $t_{f}$ is $t_0 + \operatorname{TTL}$,  and $\operatorname{TTL}$ is the time to live of the message.
For each message $m \in M$, there exists a set of nodes, $N_{m}$, that have a copy of message $m$ and a set of nodes $L_{m} = n-N_{m}$ that do not have the message.
At each distinct time, $t$ = [$t_{0}, t_{1},..., t_{f}$], encounters occur between nodes and through those encounters, messages in $M$ are transmitted.
Once a node in $N_{m}$ encounters a node, $l$, in $L_{m}$, the corresponding message, $m$, is transmitted and $l$ becomes a member of set $N_{m}$.

After $l$ obtains the message and is added to the node set, if $l \notin D$, then nothing else happens.
However, if $l \in D$, the message is considered to be delivered and the difference between the start time and the time, $z_{m} = t - t_{0_m}$, at which it occurs is the delivery latency for that message, $m$.
At each consecutive time step, more encounters occur.
Finally, when all messages are either delivered, all message $\operatorname{TTL}$'s expire, or the simulation ends; the delivery probability can be calculated as the number of messages in $M$ transmitted to $D$.
\begin{equation}
    p_{\operatorname{delivery}} = \dfrac{|M_D|}{|M|}
\end{equation}
Additionally, the upper-bound delivery latency for all delivered messages is defined as the message with the largest $z$ or:
\begin{equation}
    z_{\max} = \max(z_m)
\end{equation}

\begin{table}[htbp] 
\centering
\caption{Transition Matrices Derived from ATUS Data.
}
\label{transition_table}
\resizebox{0.48\textwidth}{!}{%
\begin{tabular}{ccc}
\toprule[1.5pt]
Time (Period) & \multicolumn{2}{c}{ Initial Probability Vector and Transition Matrices ($Home$, $Work$, $POI$)}\\
\\
\midrule
 & \multicolumn{1}{c}{Node Classification: \{$C, U, A$\}} & \multicolumn{1}{c}{Node Classification: \{$E, S$\}}\\

\\
19:00 - 06:30 (1) & (0.85, 0, 0.015)\ $
\begin{pmatrix}
0.94 & 0 & 0.064\\
0 & 1 & 0\\
0.37 & 0 & 0.63\\
\end{pmatrix}
$ & (0.70, 0.079, 0.22)\ $
\begin{pmatrix}
0.85 & 0.019 & 0.13\\
0.14 & 0.81 & 0.043\\
0.39 & 0.32 & 0.58\\
\end{pmatrix}
$\\
\\
\hline
\\
\\
06:30 - 09:30 (2) & (0.93, 0, 0.070)\ $
\begin{pmatrix}
0.97 & 0 & 0.032\\
0 & 1 & 0\\
0.59 & 0 & 0.41\\
\end{pmatrix}
$ & (0.71, 0.16, 0.13) $
\begin{pmatrix}
0.86 & 0.079 & 0.061\\
0.17 & 0.61& 0.21\\
0.51 & 0.18 & 0.31\\
\end{pmatrix}
$ \\
\\
\hline
\\
\\
09:30-16:30 (3) & (0.76, 0, 0.24) $
\begin{pmatrix}
0.89 & 0 & 0.11\\
0 & 1 & 0\\
0.36 & 0 & 0.64\\
\end{pmatrix}
$ & (0.50, 0.33, 0.13) $
\begin{pmatrix}
0.80 & 0.083 & 0.12\\
0.063 & 0.90 & 0.037\\
0.30 & 0.057 & 0.64\\
\end{pmatrix}
$\\
\\
\hline
\\
\\
16:30-19:00 (4) & (0.77, 0, 0.23) $
\begin{pmatrix}
0.91 & 0 & 0.086\\
0 & 1 & 0\\
0.30 & 0 & 0.70\\
\end{pmatrix}
$ & (0.48, 0.20, 0.32) $
\begin{pmatrix}
0.80 & 0.027 & 0.17 \\
0.042 & 0.88 & 0.78 \\
0.28 & 0.058 & 0.66 \\
\end{pmatrix}
$\\
\\
\bottomrule[1.5pt]
\end{tabular}}
\end{table}

\subsection{Mobility and transmission}\label{mobil}
The mobility of nodes in the network is described by $x$ discrete time Markov chains (DTMCs) with a finite number of states~\cite{picu2012analysis}. 
When DTMCs are used with heterogeneous contact rates, they have been shown to approximate realistic mobility for DTN scenarios and scales well with network size~\cite{picu2012analysis}.
For simplicity, the following states are used: home, work, and POI. 
Individual home and work locations are assigned to each node and POIs can be randomly selected from the set, $P$, of POIs during each transition.

The subset \{D,P\} are considered stationary nodes and do not have a transition matrix associated with them. Each mobile node in the subset $\{A,R\}$ has a unique transition matrix for each time period, $T(k) = \{t_{i}, t_{i+1}, ..., t_{\gamma}\} \mid 0 < k \leq x$. Where, each period $T(k)$ starts at $t_{i}$ and consists of $t_{\gamma} - t_{i}$ consecutive time steps.
For example, employed nodes such as $e \in E$ are preferentially attached to and are stationary at work locations, which consists of POIs in the grid during the work period (e.g. 9:30am - 4:30pm).
The model assumes that contact occurs when two nodes with the same radio are within transmission range of each other where, the transmission range is assumed to be circular.
Messages are also assumed to be small enough to be successfully transmitted within each encounter and uniformly sized.

\section{Numerical Evaluation}\label{section:evaluation}
In evaluating the feasibility and viability of the model, data was obtained from the Federal Communications Commission and the US Census Bureau regarding Owingsville, KY (Bath County)~\cite{census_quickfacts}.
Distress monitoring for cancer patients was used as a domain example of RRPM due to its delay tolerant nature (messages are valid for over 24 hours) and the prevalence of cancer and distress within eastern Kentucky~\cite{services2017,max2019opportunistic}.

\begin{table}[htbp]
\centering
\caption{Parameters used in Simulation}
\label{sim_parameters}
\resizebox{0.48\textwidth}{!}{%
\begin{tabular}{|l|c|c|}
\hline
\multicolumn{1}{|l|}{\textbf{Parameter}} & \multicolumn{1}{c|}{\textbf{Value}} & \multicolumn{1}{c|}{\textbf{Source}} \\ \hline
Simulation seeds & 0:1:99 &  --\\ \hline
Simulation duration & 24 hours & --\\ \hline
Adult Population of Owingsville & 400 & \cite{census_quickfacts}  \\ \hline
Area of Owingsville & 2.409 sqmi & \cite{census_quickfacts}  \\ \hline
Number of Cells & $820*820 = 672,400$ & \cite{census_quickfacts}  \\ \hline
Cell size & 10 ft $\times$ 10ft &  \cite{census_quickfacts} \\ \hline 
Number of patients (\( |A| \)) & 2:2:10 &  \cite{state_profiles} \\ \hline
Number of caregivers (\( |C| \)) & 2:2:10  & \cite{berry2016supporting} \\ \hline
Number of destinations (\( |D|\)) & 1  & \cite{berry2016supporting} \\ \hline
Ratio of population involved in intermediary network (\( I \)) & 0.1:0.1:1  & -- \\   \hline
Number of POIs (\( |P| \)) & 25  & Owingsville Map \\ \hline
Number of Clinical Staff (\( |S| \)) & \( \leq 2 \)  & \cite{regulations_2014} \\ \hline
Periods & 1 to 4  & \cite{hofferth_flood_sobek} \\ \hline
Data generation rate & 1 message per 24 hours  & Markey Cancer Center \\ \hline
Ratio of employed nodes & 0.935  & \cite{labor_statistics} \\ \hline
Transmission range (based on Bluetooth 5) & \( \mu = 60,\ \sigma^2 = 20 \)  & \cite{bluetooth_technology} \\ \hline
Time to live ($\operatorname{TTL}$) of a message & $\operatorname{TTL} = 24 hours$ & -- \\ \hline
\end{tabular}}
\end{table}

\subsection{Rural Node Mobility}\label{subsection:state_transitions}
Contact rates in this work are based on a 2017 IPUMS ATUS sample of non-metropolitan households in the US, 303 routine activities were obtained, along with corresponding start and stop times, and classified into three states: Home, Work and POI~\cite{hofferth_flood_sobek}.
In addition, information from IPUMS ATUS was obtained for the number of individuals in each state for 30 minute intervals, and four (4) periods were defined based on the number of people in each state.
The four (4) periods defined can be found in the "Time (Period)" column in Table~\ref{transition_table}.
Consequently, the transition matrix was estimated for each period by obtaining the transition matrix for each individual, and aggregating it over each period.

\begin{figure}[htbp]
\centering
    
  \subfloat[Mean Delivery]{\includegraphics[width=0.24\textwidth,height=1.45in]{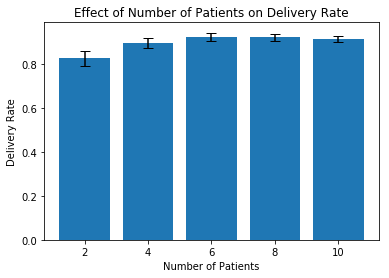}\label{Delivery_pats}}
    \subfloat[Mean Delivery Latency*]{\includegraphics[width=0.24\textwidth,height=1.45in]{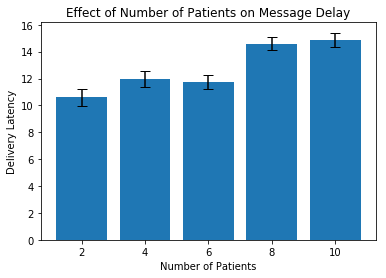}\label{Delay_pats}}
    
    \subfloat[Mean Delivery]{\includegraphics[width=0.24\textwidth,height=1.45in]{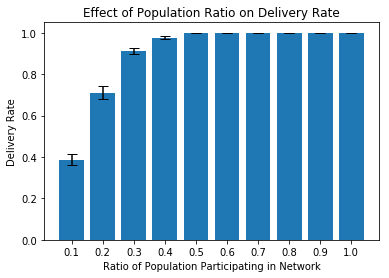}\label{Delivery_ratio}}
    \subfloat[Mean Delivery Latency*]{\includegraphics[width=0.24\textwidth,height=1.45in]{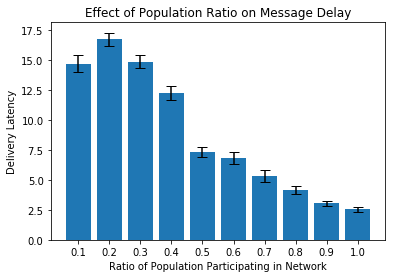}\label{Delay_ratio}}
\caption{(a) and (b) vary the number of patients in the network with .30 of the population participating in the network. (c) and (d) vary population participation in the network with 10 patients generating one message each with a time to live of 24 hours. Error bars represent SEM. * shows the delay associated with delivered messages.}
\label{results}
\end{figure}

\subsection{Simulation setup}
%-----------------------------------------------------------------------------
To understand how a DTN could be used for RRPM, distress information from cancer patients (source 
nodes) to their respective healthcare providers (destination nodes) was used 
and a simulation environment was created in Python\footnote{Code available at \url{https://github.com/netreconlab/TESCA19}}.
Table~\ref{sim_parameters} describes the parameters used in the 
simulation along with the sources for their values where applicable.

\subsection{Results}\label{subsection:results}
%-----------------------------------------------------------------------------
Figures~\ref{Delivery_pats} and \ref{Delay_pats}, give insight on the scalability of each network configuration when 30\% of the population participates in an intermediary network.
The results show that an optimal delivery rates of 38.7\% -- 100\% can be achieved within message delivery latencies of $\sim$4 and $\sim$17 hours.
Further, one sees that increasing the number of intermediary nodes exponentially reduces the delay while increasing the number of patients linearly increases the message delivery latency.

%-----------------------------------------------------------------------------
\section{Discussion}\label{section:discussion}
The authors evaluate an architecture that transmits patients' health data opportunistically 
until it reaches healthcare providers. 
The simulation results, using real-world data from Owingsville, KY, a small 
rural Appalachian city, have demonstrated that the proposed model is feasible 
and can provide a timely and reliable communication to remotely link rural patients with 
their providers; resulting in better quality of care. 
The findings show that with only 0.30 rural adult population participation, the architecture can deliver 0.91 of non-emergency medical information with an average delivery latency of $\sim$13 hours. 
This indicates that the architecture can work for domains in which monitoring is done on daily basis such as remotely acquired Patient Reported Outcome Measures.
The network design analyzed in this study can only be used for non-time critical data.

\bibliographystyle{ACM-Reference-Format}
\bibliography{tesca}

\end{document}